\documentclass[useAMS,usenatbib]{mn2e}
\usepackage{graphicx}
\usepackage{natbib}

\title[Field Configuration by Simulated Annealing]{Multi-Object Spectroscopy Field Configuration by Simulated Annealing}
\author[Brent Miszalski et al.]
{
Brent Miszalski$^{1,2}$\thanks{E-mail: brent@ics.mq.edu.au}, K. Shortridge$^{2}$, W. Saunders$^{2}$, Q.A. Parker$^{1,2}$ and S.M. Croom$^{2}$ \\
$^{1}$Department of Physics, Macquarie University, Sydney, NSW 2109, Australia,\\
$^{2}$Anglo-Australian Observatory, Epping, NSW 1710, Australia
}
\date{ Accepted 2006 June 30.  Received 2006 June 30; in original form 2006 February 17}

\begin{document}
\maketitle
\begin{abstract}
Multi-object spectroscopy (MOS) instruments, such as the Two-degree Field (2dF) facility of the Anglo-Australian Observatory (AAO), have facilitated large-scale redshift surveys. Yet despite their acclaim, instrument design has been suspected of introducing subtle selection effects into surveys. Investigation into these selection effects has been overshadowed by instrument complexity. We identify the field configuration algorithm (FCA) used to select targets for observation as mainly responsible for such effects. A FCA can imprint artificial structure on observed target distributions, which may accrue over large angular scales, potentially to the detriment of statistical analyses applied to such surveys.
We present here a new FCA developed for 2dF that is based on simulated annealing (SA), a generic method commonly used to solve constrained optimisation problems. We generate synthetic fields and utilise mock 2dF volumes to contrast the behaviour of previous strategies with the SA FCA. The angular two-point correlation function and other sensitive techniques reveal that the new FCA achieves unprecedented sampling uniformity and target yield with improved target priority handling and observational flexibility over current FCAs. The SA FCA is generic enough to be used by current 2dF-like and potentially next-generation MOS instruments with little modification.
\end{abstract}

\section{Introduction}
The scientific motivation for large-scale redshift surveys has driven the development of efficient multi-object spectroscopy (MOS) instrumentation such as the 2dF facility (Lewis et al. 2002). Their high multiplex advantage is typically achieved by the placement of optical fibres in the focal plane of a telescope to relay light from multiple astronomical targets to a spectrograph. However, this placement can be severely constrained by the physical design of the instrument. Such constraints pose significant challenges to a field configuration algorithm (FCA) if it is to uniformly sample targets for observation whilst upholding the multiplex advantage of the instrument. Uniform sampling is essential to minimise any artificial power imprinted by a FCA on survey target distributions that are later subjected to sensitive statistical analyses.

Although robotic fibre placement systems offer greater efficiency than manually operated counterparts, they are undoubtedly the most complex (for an overview see Smith et al. (2004)). Even the most basic of operations, such as observing a field, consists of multiple stages. Before a field can be observed a FCA is used to create a mapping between fibres and targets. This mapping is then used to determine a set of fibre movements that the positioner must make before the field can be exposed. The significant engineering effort required to manipulate fibres (e.g. Wilcox 1993) has overshadowed the requirements of field configuration, leaving FCAs relatively underdeveloped.

Preliminary FCAs made possible the instrument design studies of Donnelly et al. (1992; hereafter DON92) and Lewis et al. (1993; hereafter LEW93). These works helped shape the final design of their respective instruments by optimising parameters such as button shape, angular fibre deviation and field geometry. Additionally, DON92 investigated two disparate field configuration strategies. The ideal strategy was an `exhaustive' approach, whereby possible solutions were explored by extensive field randomisation guided by criteria that describe an optimal configuration. DON92 found that although an `exhaustive' strategy would generate very optimal configurations, its heavy use of computer resources meant that a more `intelligent' approach was more viable at the time. The `intelligent' strategy mimics a very clever and patient human who iteratively recognises optimal moves towards a final configuration. Such `intelligent' strategies were readily adopted as the basis for early FCAs because of their relative speed and their ability to generate solutions of comparable quality to `exhaustive' methods. 

A major disadvantage of `intelligent' FCAs was their strong dependence on understanding how field plate components interact. The emergence of MOS in the early nineties meant that this understanding was not sufficiently mature enough to enhance the development of early FCAs. Substantial progress was later made coinciding with the 2dF Galaxy Redshift Survey (2dFGRS; Colless et al. 2001, hereafter COL01), which utilised the `Oxford' FCA that was tailor-made to 2dFGRS fields (\S5.1; COL01). The `Oxford' FCA used an insight into fibre availability to attempt uniform target sampling of 2dFGRS fields. Furthermore, its use of fibre swaps to steadily optimise a configuration resulted in much higher target yields. Indeed, much of the success of the 2dFGRS can be attributed to the `Oxford' FCA, which has long been the default FCA for the 2dF \textsc{configure} program that has been used to prepare fields for observation.

Despite such progress, the effect FCAs have on target sampling in large-scale redshift surveys has remained an open question. Such an influence has long been thought to exist, albeit at a negligibly small level. Unless a finished survey has very high completeness, the FCA is likely to have some measurable influence. Although great scrutiny of the source and propagation of selection effects in the construction of such surveys is standard practice (e.g. colour and magnitude selection; tiling algorithm), attempts to measure FCA influence are somewhat of an afterthought. Such attempts are often left until a survey is complete where the perceptible influence of a FCA appears relatively benign. This approach bypasses detailed understanding of FCA behaviour, independent of any tiling algorithm, before a survey is designed. This preference has been exacerbated by a lack of sensitive analytical tools to quantify these influences.

The preliminary work of Outram (2004; hereafter OUT04) addressed these issues by implementing sufficiently sensitive analytical tools and using them in a systematic fashion. Under certain conditions, OUT04 discovered previously unknown artificial structure imprinted by the `Oxford' FCA, highlighting the need for precaution when designing large-scale surveys. The work of OUT04 was expanded upon by Miszalski (2005; hereafter MIS05), in addition to implementing a new FCA based on simulated annealing (SA), an `exhaustive' method, that exploits the considerable computer power now available to produce highly optimal field configurations. MIS05 used a wider variety of synthetic fields than OUT04 to contrast the `Oxford' and SA FCAs, concluding that the `Oxford' FCA was relatively unsuitable for the needs of 2dF after the AAOmega spectrograph upgrade (Saunders et al. 2004). 

This paper serves a dual purpose to (i) Review field configuration requirements and strategies within the context of 2dF-AAOmega (Section 2 and throughout) and (ii) Describe the SA FCA developed by MIS05 for 2dF-AAOmega in Section 3. Section 4 describes the synthetic fields generated to facilitate a comparative study of the `Oxford' and SA FCAs in Section 5. Section 6 concludes with a summary of the SA FCA and its performance.

\section{Field Configuration}
\subsection{Problem Description}
Field configuration is the task of selecting astronomical targets for observation with MOS instruments. The most basic objective of FCAs is to maximise the allocated target yield to ensure that the multiplex advantage of the MOS instrument is upheld. Field configuration is hindered by many instrumental and observational constraints that we introduce here in the context of 2dF-AAOmega. 

\begin{figure}
        \begin{center}
                \includegraphics[scale=0.6]{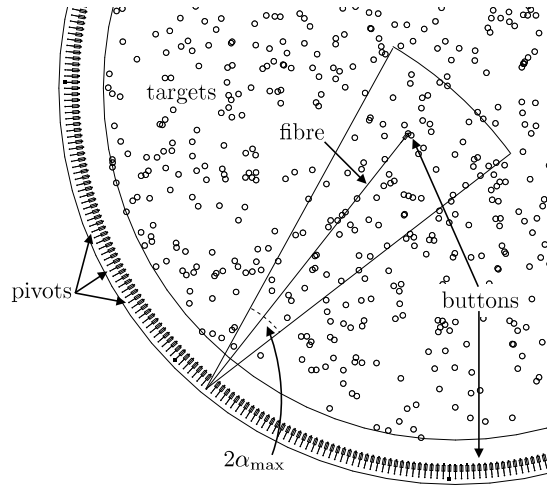}
        \end{center}
        \caption{A schematic of part of a 2dF field plate with key components involved in field configuration labelled. An allocation takes place if the button of an optical fibre coincides with the position of a target projected on the focal surface. The placement of each fibre is restricted within a sector (outlined) that subtends an angle 2$\alpha_\mathrm{max}$, where $\alpha$ is the angular fibre deviation. A pivot is a point from which fibres can be drawn, a term that is used interchangeably with fibre.}
        \label{fig:plate}
\end{figure}

Figure \ref{fig:plate} depicts the key components of 2dF field configuration. A target selected for observation must be allocated to a fibre, whereby a fibre's magnetic button is robotically placed such that its microprism is precisely aligned with the projected image of the target on the focal surface. A field consists of 400 fibres that can be drawn from 400 (\texttt{NPiv}) evenly spaced pivot points around the field plate perimeter. Each field contains a user-specified number of targets (\texttt{NTargets}) within the perimeter, and we define $\tau$ as the ratio \texttt{NTargets}/\texttt{NPiv}, where \texttt{NPiv} may very depending on the presence of broken fibres. However, not all targets are accessible to a fibre. A fibre may only allocate targets within a sector (Figure \ref{fig:plate}) that the fibre sweeps out because fibres can only be extended to 1.12$R$, where $R$ is the field radius, and have a maximum angular deviation of $\sim$14 degrees from the radial direction. This means a single fibre can cover $\sim$8.7 per cent of the field plate area. We assume an artificial limit of $\sim$14 degrees for all fibres that restricts the range of fibre deviation to help increase the range of hour angle over which allocations are valid. 

\begin{figure}
        \begin{center}
                \includegraphics[scale=0.35,angle=270]{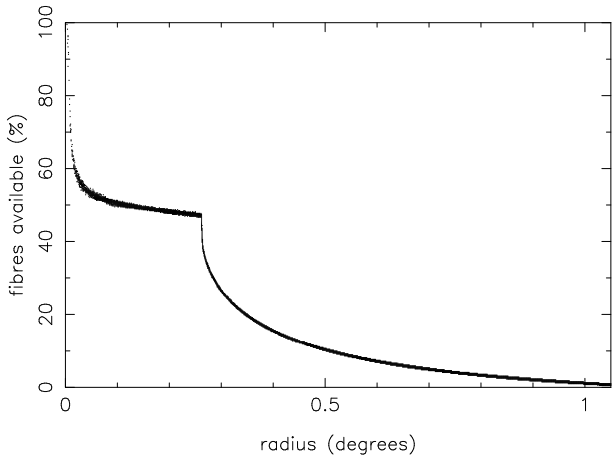}
        \end{center}
        \caption{The empirically determined percentage of fibres that can access targets at a given radial distance from the field centre for 2dF-AAOmega. The first tier arises as finite fibre length is exceeded for half the fibres, the second is attributable to fibre deviation reducing near the field plate edge. A strong inherent bias towards central targets is evident. This bias has been particularly difficult to address with FCAs primarily because of the paucity of fibres available past $r$$\sim$0\fdg5.}
        \label{fig:reach}
\end{figure}

Figure \ref{fig:reach} depicts an instrumental bias towards central targets empirically determined using high target density fields in the absence of a FCA. The first tier arises as finite fibre length is exceeded for half the fibres. The $\sim$0.5 degree diameter of this bias roughly corresponds to the chord length of the sector traced out by a fibre. The second tier is attributable to fibre deviation reducing near the field plate edge. On smaller angular scales target separation is primarily constrained to $\sim$30 arcsec by the physical footprint of fibre buttons, with secondary constraints arising from the effect of button-fibre tolerances up to separations of $\sim$2.5 arcmin. These constraints are somewhat alleviated by the ability to cross fibres, a feature that extends field configuration flexibility and increases target yields.

There are three classes of targets observable with 2dF. Program (P) targets are those of principal scientific interest, e.g. galaxies. Fiducial (F) targets are stars of reliable position used for field acquisition and guidance. Sky (S) targets are blank-sky positions which provide night sky background for subtraction from program target spectra. Program and sky targets are allocated via 392 science fibres, while 8 dedicated fibre bundles are deployed from the cardinal and primary inter-cardinal points on the field plate perimeter to fiducial targets.

The method of sky target allocation strongly determines the accuracy of sky subtraction from program target spectra. Saunders (2005) details three methods available with 2dF-AAOmega. The standard technique involves allocating 1.25$\sqrt{\texttt{NPiv}}$ science fibres to sky targets to achieve an accuracy of $\sim$1--2 per cent. Alternatively, the nod and shuffle (N\&S; Glazebrook \& Bland-Hawthorn 2001) technique can be used to increase this accuracy to $\sim$0.1 per cent.  In N\&S the telescope nods between program targets and an offset sky field whilst shuffling rows on the spectrograph CCD. With AAOmega an improvement on N\&S, called N\&S with cross-beam switching (hereafter CBS), can be used to reduce the signal to noise penalty from 2 per target for N\&S to $\sqrt{2}$. CBS involves the continuous observation of individual program targets with two closely separated fibres. The telescope nods between each pair of fibres receiving alternate sky and program spectra, enabling very high spectral density to be achieved. Maximising CBS pair yield is particularly challenging for FCAs. Only 98 CBS pairs are observable as each AAOmega CCD has limited charge-shuffling area. The fibres allocated to pairs need not be adjacent on the fibre slit as the spectra are co-added post-observation.

Initially target priorities were used solely to counteract the geometrical bias depicted in Figure \ref{fig:reach} (e.g LEW93). A common feature of modern FCAs, such as those used for 2dF field configuration, is the provision for nine different target priorities from 1 (lowest) through 9 (highest). These are used by observers to give preference to more scientifically valuable targets. It is common for large surveys, such as the 2dFGRS, to assign different target priorities to separate target populations such as quasi-stellar objects (QSOs) and galaxies. COL01 describes how target priorities are assigned primarily from tiling considerations and later incremented by one if the target is a QSO. The special treatment of QSOs attempts to be a proactive measure to prevent the imprint of the galaxy clustering signature on the weaker clustered QSOs. This is despite the lack of priority-specific analysis into the effect FCAs have on the sampling of targets.

\subsection{Finding an Optimal Solution}
A FCA endeavours to obtain an optimal field configuration under many instrumental and observational constraints. We characterise an optimal solution to field configuration by the following optimality criteria:
\begin{enumerate}
        \item High overall yield independent of target priority
        \item Highest priority targets have highest possible yields
        \item Uniform sampling overall and for each priority
        \item Observational flexibility
\end{enumerate}
Criterion (i) is tantamount to upholding the multiplex advantage. Target priorities are weighted correctly if (ii) is satisfied so that they can be reliably used however observers see fit. Minimal detrimental artificial power is introduced into allocated target distributions if (iii) is met. A corollary of (iii) is that the field-of-view of the instrument not be diminished by geometric biases. A flexible approach to sky subtraction techniques and different fields and their target distributions is ensured if (iv) is met. We omit time variable aspects of field configuration, such as hour angle effects and broken fibres, from (iv) since these are typically handled independently within the FCA software infrastructure. If such events were to occur then it would be a simple matter of reconfiguring the field with some time penalty.

The optimality criteria are not mutually exclusive. For instance, it would be undesirable to have sky targets allocated at the expense of high priority targets when lower priority targets remain allocated. This complicates the development of a FCA that satisfies all the criteria. Furthermore, such considerations have to take place within the instrumental and observational constraints of field configuration that classify it as a constrained optimisation (CO) problem.

CO problems aim to maximise an objective function $f$ that describes the quality of a system being optimised, subject to the constraints specified by the function $g$. The parameter space $S$ is the set of all possible solutions to the problem. In many problems of interest $f$ and $g$ are non-linear functions of very many variables. This makes CO problems very difficult to solve in general, which is often compounded by $S$ being factorially large. If a solution is found, then it is very unlikely that it corresponds to a global maximum, i.e. an optimal solution.

Previous FCAs have utilised the `intelligent' strategy of DON92, whereby $S$ is explored in an `intelligent' fashion, within a framework akin to conjugate gradient methods (e.g. Press et al. 1992). Specifically, they navigate $S$ by climbing local maxima until no further improvements can be made. Such `intelligent' methods depend critically on precise knowledge of field configuration in order to mimic the iterative moves that a human operator would make towards an optimal solution. If this knowledge is ill-defined or cannot be simply addressed by the FCA, then this can skew the search within $S$ towards undesirable, sub-optimal solutions. For instance, if an assumption that fibre crossovers are to be avoided is made, then an optimal solution would have minimal crossovers, but poor target yield because crossovers are beneficial in this regard.

However, if the problem is reasonably well-defined and catered for by the FCA, then solutions of comparable quality to those obtained from an `exhaustive' strategy can be obtained, often with fewer computer resources and in less time. We discuss the implementation of a new `exhaustive' FCA in the next section.

\section{Field Configuration Algorithm}
\subsection{Simulated Annealing}
A radically different approach to the `intelligent' FCA paradigm is to search the parameter space $S$ in an `exhaustive' fashion with a random walk. This approach requires almost no insight into the problem at all, save for a well-defined objective function $f$ that constrains the random walk in the parameter space. DON92 first proposed simulated annealing (SA; Kirkpatrick et al. 1983) as a suitable algorithmic framework for such a FCA, however it was not viable at that time. SA is a resource intensive prescription for solving CO problems with a simple structure. The resource intensive nature is a caveat regarding the lack of required insight into the problem. Fortunately, we have recently found modern computing power to be viable to configure fields with SA.

SA employs the probabilistic Metropolis algorithm (Metropolis et al. 1953) to simulate the thermal motion of a system to be optimised (in this case a field) whilst slowly cooling the system to simulate the annealing process of physical systems (e.g. metals). The simulation introduces small, random perturbations (fibre swaps) into the system at each stage in the cooling process, known as the annealing schedule, that, when guided by the Metropolis algorithm, enable an effective search of $S$ that frequently leads the algorithm towards an optimal solution. Indeed, the strong dependence on $f$ and its maximisation by the Metropolis algorithm makes $f$ a simple and powerful interface for encapsulating the optimality criteria. This enables the criteria to be satisfied in a maintainable and seemingly effortless fashion.

The choice to use SA over neural network or genetic algorithms is justified by two precedents. The first is the aforementioned implementation of a SA FCA by DON92. Secondly, Campbell et al. (2004) used SA to solve the tiling problem for astronomical surveys, the closest problem to field configuration. Field configuration inherits observational constraints from the tiling problem, specifically the priority and spatial distribution of targets. In this respect the most thorough tiling algorithms would incorporate field configuration to ensure that post-tiling target attrition arising from instrumental constraints is minimised. 

The remainder of this section describes the infrastructure within which the SA FCA was developed followed by the algorithm itself.

\subsection{Software Infrastructure}
The field configuration process for 2dF is performed by the \textsc{configure} program (Lewis et al. 2002). \textsc{configure} includes many FCAs that can be chosen to prepare user provided target coordinate lists that constitute a field (an \textsc{ascii} .fld file) for observation with 2dF. The default output of \textsc{configure} contains data on target coordinates, allocation statuses and priorities. These data are typically used by the 2dF control software to prepare a field for observation but the data from hundreds of individual fields can also be collated later to quantify the extent to which an FCA meets the optimality criteria.

We use a batch version of \textsc{configure} that serves as our development platform for the new FCA. This version of \textsc{configure} inherits the design of \textsc{configure} and its generic interface for algorithm development that is largely independent of instrument design. This is facilitated by many different instrument profiles that handle the associated different field geometries and constraints. The SA FCA was developed specifically for the 2dF-AAOmega instrument profile, however it can also be used by other instrument profiles. 

\subsection{Pre-Configuration}
The pre-configuration stage is responsible for establishing the parameter space $S$ of a field for a FCA to explore. Although this task is largely dealt with by existing \textsc{configure} code, preexisting collision detection strategies were inadequate for our purposes. Such strategies are necessary to determine the validity of allocations proposed by a FCA. Traditionally \textsc{configure} invokes collision detection routines afresh for each proposed allocation, creating a large computational overhead. This slows down FCAs and limits the extent of $S$ they can explore. 

The requirement of intensive field randomisation, which we describe later as fibre swaps, has motivated the development of the allocation sub-system built into the SA FCA. It pre-calculates all possible conflicts between all possible allocations for a field and stores them in a sparse, indexed collision matrix. An efficient process of elimination is employed to generate the matrix from simple determinations (e.g. fibre reach) to progressively more geometrically complex ones (e.g. button-fibre proximity). The FCA can then perform a fast table look-up to determine whether a proposed allocation is valid. This removal of the collision detection bottleneck reduces the typical annealing schedule run time from over one hour to just a few minutes on a reasonably powerful modern computer. The lack of large amounts of physical memory, and the fast processing power to fill it, has hitherto precluded such techniques from FCA design. 

However, there are a number of limitations to the allocation sub-system. The most significant limitation is the calculation of the collision matrix, which typically takes $\sim$5 minutes for fields of moderate target density ($\tau$$\sim$1). Once calculated, the matrix can be reused to configure the field with different algorithm parameters. The matrix is valid until the spatial distribution of targets changes, that is when targets are introduced or removed, when the field centre changes or the hour angle becomes invalid. The time required to calculate the matrix means field configuration with the SA FCA is best classified as semi-interactive. For the majority of moderate density fields 1GB of physical memory suffices.

Unfortunately, not all fields are amenable to the construction of an associated collision matrix, and thus field configuration by the SA FCA. It is assumed that most observers will be diligent enough not to include excessive target numbers (e.g. 1000 or more) nor to `pack' targets too close together into a small region of the field plate. After all, 2dF only has 400 fibres that under-sample such large target numbers and can struggle with close targets because of the physical footprint of field plate components. Naturally, one can expect the best performance from an FCA when a field is well matched to instrument specifications. However, there is some capacity within the current allocation sub-system to accommodate moderately dense fields, provided sufficient physical memory is available (2GB recommended). Unfortunately, such fields will involve considerably longer matrix calculation times than `standard' fields well matched to the 2dF specifications. 

If collision matrix calculation for a field is problematic, then the observer should strongly consider sparse-sampling the field. Currently, tedious sparse-sampling is the only possible course of action for these fields. Alternative methods are currently under investigation. One approach involves the random exclusion of fibres prior to collision matrix calculation. Fibres are selected for exclusion if they are amongst many fibres that can reach a target or if they can reach a low priority target. This normalises the matrix setup time by reducing the number of possible allocations until it reaches an acceptable level. It is uncertain at this time the effect to which such fibre exclusion will have on target distributions.

\subsection{The Annealing Schedule}
An outer loop simulates the cooling process with an initial temperature $T_i$ established. At each iteration the temperature $T$ is reduced by multiplication with $1-\Delta T$, where $\Delta T$ is small, to slowly cool the system. The annealing schedule exits when either the final temperature $T_f$ is reached or the user interrupts the configuration. The temperature is a control parameter of the same units as the objective function (see later).

A traversal over all pivots to randomly select new targets for each fibre constitutes the core of the annealing schedule. This occurs at each temperature stage and can be repeated \texttt{MaxIters} times to ensure that the system quasi-statically reaches thermal equilibrium. An upper bound to the number of swaps per temperature stage over the whole field is given by \texttt{MaxIters} times \texttt{NPiv}, which is $\sim$10$^5$ swaps for default parameters.

\begin{figure}
        \begin{center}
                \includegraphics[scale=0.4]{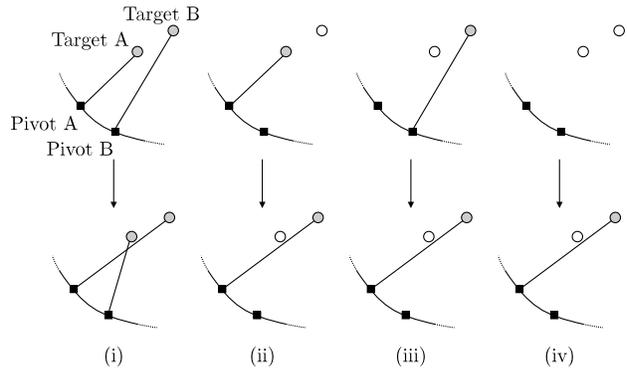}
        \end{center}
        \caption{The four randomisation cases possible when given \texttt{PivotA} and a randomly selected target \texttt{TargetB}. The type of randomisation is determined by the possibility that \texttt{PivotA} may be allocated to \texttt{TargetA} and \texttt{TargetB} may be allocated by \texttt{PivotB}. The initial configuration (top) is changed which results in the final randomised state (bottom).}
        \label{fig:random}
\end{figure}

Randomisation of a field is performed one pivot at a time during the pivot traversal. The algorithm randomly selects a new target (\texttt{TargetB}) for each pivot (\texttt{PivotA}) from a list of targets that the pivot's fibre can physically reach. Possible courses of action at this point were the subject of considerable experimentation. We identified a set of four different paths (Figure \ref{fig:random}) that can be taken depending on the allocation status of \texttt{TargetB} or \texttt{PivotA}. The general aim of each path is to allocate \texttt{PivotA} to \texttt{TargetB} whilst conserving target yield. 

Each path works cooperatively with the others. If some paths are excluded, then insufficient field randomisation may occur. Although different paths are possible, those shown here provide sufficient randomisation without restricting the speed of the algorithm significantly. The introduction of `pseudo-deallocations' by the `swapping-off' action of type (ii) swaps can give type (iv) swaps the chance to clean up any remaining unallocated targets. Indeed, the efficacy of this behaviour has meant that no deliberate fibre deallocations are necessary to randomise the field. 

The randomisation of each fibre during the annealing schedule is accepted only if the Metropolis algorithm is satisfied. A randomisation is accepted with probability determined by
\begin{equation}
   \label{eqn:metro}
P = \left\{ \begin{array}{ll}
	1 & E_2 \ge E_1 \\
	e^{\Delta E/T} & E_2 < E_1,\\
\end{array}\right.
\end{equation}
where $\Delta E = E_2 - E_1$ and $e^{\Delta E/T}$ is only accepted if $\xi < e^{\Delta E/T}$, where $\xi \in [0,1]$ is a random number. Traditionally if $E_2 < E_1$ the operation would not occur, but the Metropolis algorithm ensures a non-zero possibility of acceptance that provides the algorithm with the ability to avoid local maxima. Here $E$ is the energy or quality of the field as determined by the objective function 
\begin{equation}
	\label{eqn:objective}
	E = \sum_i^\mathrm{NPiv} \left[ \beta^{p_i} + \delta \sum_j^\mathrm{NAssoc} \beta^{p_j} \right]\left(\frac{\alpha_i - \alpha_\mathrm{max}}{\alpha_\mathrm{max}}\right)^\gamma ,
\end{equation}
where $\beta$, $\gamma$ and $\delta$ are real parameters, $\alpha_{\mathrm{max}}$ is the maximum angular fibre deviation, and $\alpha_i$, $p_i$ represent the angular fibre deviation and priority of the allocation to pivot $i$. The first $\beta$ term is a priority weighting identical to that used by Campbell et al. (2004). The second $\beta$ term is a close-pairs constraint that favours \texttt{NAssoc} close-pairs around the target allocated to pivot $i$. This term is later used to address close-pair deficiencies, with possible application in surveys, and to maximise the number of CBS pairs. 
The remaining term is an angular fibre deviation constraint that favours straight fibres. This term may help reduce reconfiguration time between fields. Although the terms presented here are preliminary, they have been found experimentally to yield excellent results. 

No further constraints outside of the objective function on the allocation of fiducial and program targets are made, although the situation is different for sky targets. Traditionally, sky targets have been allocation after the main allocation pass in post-configuration routines (e.g. `Oxford' and Roll et al. (1998) FCAs). This involved the reservation of a set number of fibres during the main pass or simply reassigning already allocated fibres to sky targets. 

Alternatively, we include sky targets in the annealing schedule. They take part in the same fibre swaps as program and fiducial targets with some conditions. Type (ii) swaps are prohibited if \texttt{TargetA} is a sky target and \texttt{TargetB} is a program target. Furthermore, type (ii) and (iv) swaps are prohibited if \texttt{TargetB} is a sky target and the sky target quota has already been met. Although these conditions are relatively weak, the quota is typically met. If the sky target density is particularly low, then additional weight may be given to sky targets to ensure the quota is enforced.

The advantage of including sky targets in the annealing schedule is not immediately obvious. It will be shown later that this approach is preferable because it does not sacrifice higher priority targets to allocate sky targets, despite lower priority targets still available, unlike previous methods.

\subsection{Post-Configuration}
There are currently no post-configuration routines associated with the SA FCA. This can be attributed to the simple structure of SA, centred on the objective function. Sky allocation is integrated in the annealing schedule. A fibre swapping routine, that may otherwise free-up fibres, is not necessary by virtue of the high degree of randomisation the annealing schedule imparts on the field. We do not implement a fibre uncrossing routine here, primarily because of the existence of the angular fibre deviation constraint. This does not necessarily preclude future development of an uncrossing routine.

\section{Field Generation}
\subsection{Requirements}
To quantify the extent to which a FCA meets the optimality criteria requires the ability to generate large sets of synthetic fields of various types. These fields could then be configured using our batch version of \textsc{configure} and later processed to obtain detailed statistics on how FCAs treat different target distributions. As part of the testing of their FCA, DON92 generate four different types of fields with random target distributions that are:
\begin{enumerate}
        \item Uniform
        \item Weakly clustered at field centre
        \item Strongly clustered at field centre
        \item Similarly clustered to Maddox et al. (1990) galaxies
\end{enumerate}
Uniform target distributions help elucidate the general performance of FCAs in addition to isolating instrumental constraints. The second and third types replicate densely populated fields (e.g. globular clusters). The last type replicates the most common `real-world' fields of galaxy redshift surveys. These four field types form the basis for any thorough analysis of FCAs and we adopt them for this work. 

A further requirement we add is the ability to assign target priorities and to generate sky and fiducial targets. We emphasise that these are synthetic fields intended to explore the response of FCAs to different spatial and priority variations in target distributions by Monte-Carlo simulations. It is beyond the scope of this paper to consider the effect of depth in redshift space in the context of tiling algorithms for redshift surveys. 

\subsection{Uniform and Gaussian Fields}
\begin{figure}
        \begin{center}
                \includegraphics[scale=0.24]{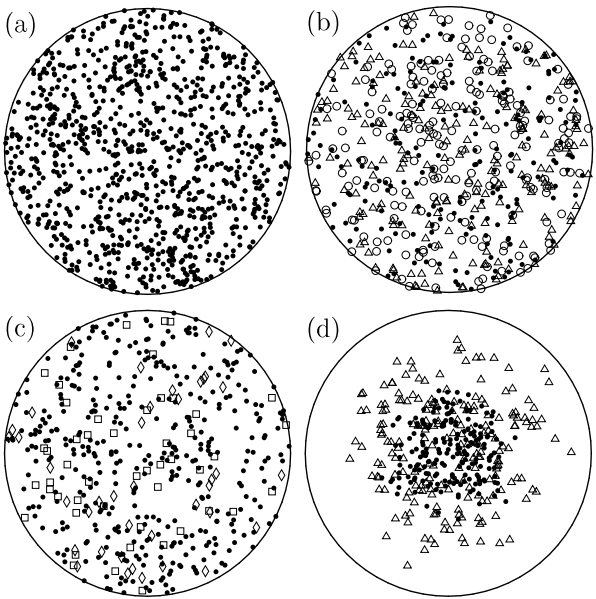}
        \end{center}
        \caption{Sample fields created with \textsc{fieldgen}. (a) Uniform distribution of 1000 targets. (b) Stacked uniform distributions of 200 targets each for priorities 7 (circles), 8 (triangles) and 9 (points). (c) Uniform distribution of 400 targets with uniform distributions of 40 sky (diamonds) and 50 fiducial (squares) targets. (d) Stacked Gaussian distributions of 200 targets each for $\sigma = 0.1$ (points) and $\sigma = 0.3$ (triangles).}
        \label{fig:fields}
\end{figure}

The first three field types are generated using our \textsc{fieldgen} program. It uses the \textsc{math::random} Perl module\footnote{Available from http://www.cpan.org} to generate both uniform and Gaussian (centrally clustered) random target distributions. Multiple target populations of specified number and priority can be stacked to form complex fields. Uniform sky and fiducial target distributions are also supported. A selection of fields generated by \textsc{fieldgen} is shown in Figure \ref{fig:fields}. Quality evaluation of the target generation code showed that the occurrence of identical targets is negligible amongst 1000 fields, each with 1000 uniformly distributed targets. Further reassurance is provided by negligible angular correlation between targets from those fields (see later).

\subsection{Clustered Fields}
To gauge the real-world performance of FCAs, fields are required that have clustered target distributions. The uniform fields generated by \textsc{fieldgen} are inadequate in this regard because the underlying target distribution is random. Clustered fields analogous to field type four are difficult to generate from first principles (e.g. Infante et al. 1994), however preexisting data can be exploited for this purpose. We make use of the mock 2dF galaxy redshift survey catalogues (Cole et al. 1998) in this work, however we are only concerned with the angular clustering of targets in this work.

The mock 2dF catalogues contain synthetic galaxies generated from large, high-resolution N-body simulations. They mimic the geometrical construction and the expected clustering properties of 2dFGRS data. The moderate clustering in redshift space provides an excellent resource for clustered field generation when projected onto the sky. We utilise the north and south galactic pole (NGP and SGP) catalogues for cluster normalised, flat cosmology data with $\Omega_0=0.3$ and $\Lambda_0=0.7$. 

\section{Results}
\subsection{Parameter Selection and Performance Evaluation}
\begin{figure}
        \begin{center}
                \includegraphics[scale=0.35,angle=270]{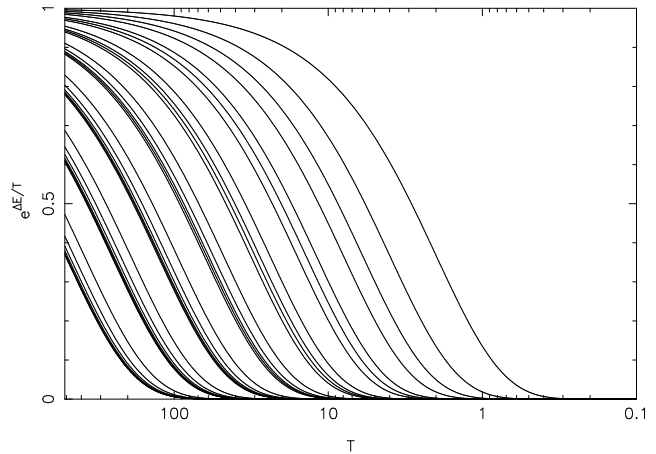}
        \end{center}
        \caption{Families of $e^{\Delta E/T}$ curves that show the probability of accepting unfavourable steps in field randomisation. Each curve is generated by permutations of $\Delta E$=$\beta^{p_f} - \beta^{p_i}$, where $\beta$=2, $p_f$ and $p_i$ are the target priorities after and before randomisation (note $p_f < p_i$). Optimal selection of $T_i$ and $T_f$ may be determined by choosing the boundary values within which curves similar to these lie, for example $T_i$=0.1 and $T_f$=512 ($\beta^9$) in this case.}
        \label{fig:params}
\end{figure}

The optimisation of even the simplest FCA is a difficult task considering the large variety in the spatial and priority distribution of targets possible within fields. This complicates the choice of the annealing schedule parameters $T_i$, $\Delta T$, $T_f$ and \texttt{MaxIters} that are critical to obtaining an optimal solution. Ideally this would be achieved by constructing families of curves in a plot of $e^{\Delta E/T}$ versus $T$ (e.g. Figure \ref{fig:params}). This gives the probability of accepting all types of unfavourable steps encountered in randomising a field as the annealing progresses. Selection of $T_i$ and $T_f$ would then proceed to roughly demarcate the families of curves to ensure that each curve, representing a particular random transition, is given adequate treatment by the Metropolis algorithm, so that one curve is not always accepted or rejected.

In reality such treatment is complex with numerous curves that are coupled to the unique parameters of the objective function. 
It turns out that poor choices of $T_i$ and $T_f$ can be offset by sufficient field randomisation which takes place if $\Delta T < 0.1$ or \texttt{MaxIters} $\ga$ 20. For these reasons we choose nominal values of $T_i$ in the range 10--512 and $T_f$=0.1 with \texttt{MaxIters}=20, $\Delta T$=0.01, $\beta$=2, $\delta$=0 and $\gamma$=0 unless otherwise stated for the following tests. 

The following tests contrast the performance of the SA FCA against the `Oxford' FCA. 
We acknowledge that although the `Oxford' FCA was highly optimised for 2dFGRS fields, this is still a worthwhile comparison, primarily because of the `Oxford' FCA's prolific use in other programs besides 2dFGRS and its inevitable use as the sole FCA for 2dF-AAOmega had another FCA not been developed. Where appropriate we state whether the `Oxford' FCA is optimised for a particular task. We use the default parameters for the `Oxford' FCA.

All fields are either generated by \textsc{fieldgen} or extracted from mock 2dF catalogues. They are configured both by the SA and `Oxford' FCAs via our batch version of \textsc{configure} using the 2dF-AAOmega instrument profile. All 392 science fibres are available for program and other fiducial or sky targets (if present). We exclude fiducial and sky targets from the analysis except where stated to isolate FCA performance on program targets. One allocation pass is performed on all fields, no reconfiguration of fields takes place. This means any gain seen in target yields of the SA over the `Oxford' FCA are an upper limit because the latter is known to `pick-up' a few additional targets upon reconfiguration. 

\subsection{Target Yield}
\subsubsection{Single Priority Fields}
Fields with $\tau$=\texttt{NTargets}/\texttt{NPiv} being $\gg 1$ do not suffer from poor target yields because of the high excess of targets available per fibre. More challenging fields with $\tau \la 1$ have traditionally been a weakness of FCAs because of limited targets to swap fibres between to increase yields. 
To determine the gain in overall target yield that the SA FCA can produce over the `Oxford' FCA, two field sets with $\tau \la 1$ were examined. 

\begin{figure}
        \begin{center}
                \includegraphics[scale=0.35,angle=270]{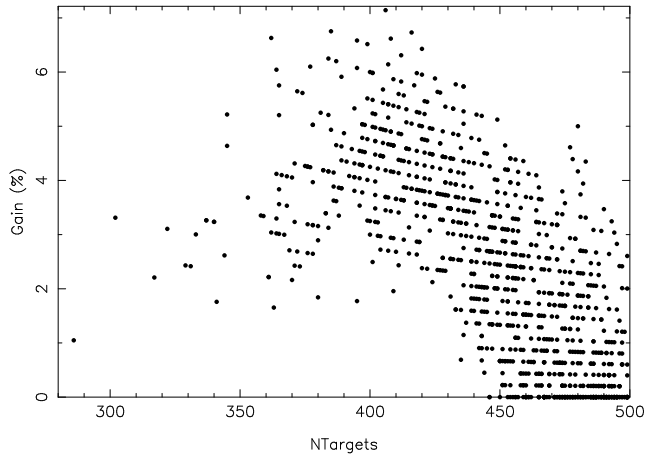}           
        \end{center}
        \caption{The gain envelope seen shows the overall target yield acquired by the SA FCA over the `Oxford' FCA for 1000 moderately clustered mock 2dF fields. The striations are a quantisation effect.}
        \label{fig:mockyield}
\end{figure}

The first set consists of 1000 randomly selected fields extracted from the mock 2dF catalogues. Each field has 446 moderately clustered priority 9 (P9) targets on average. We increased \texttt{MaxIters} to 40 and chose a moderate value of $T_i$=100 for the SA FCA to accommodate the challenging fields. In all cases the SA FCA outperforms the `Oxford' FCA by up to 7 per cent (Figure \ref{fig:mockyield}). Moderate target clustering often precludes all fibres being allocated in the low and high target number limits. 

\begin{figure}
        \begin{center}
                \includegraphics[scale=0.35,angle=270]{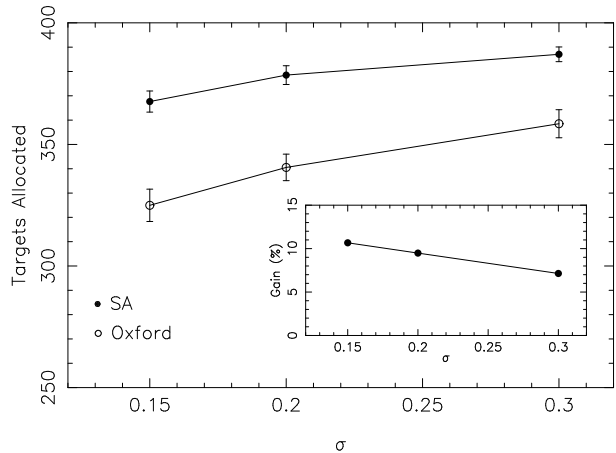}           
        \end{center}
        \caption{The overall target yield acquired by the SA and `Oxford' FCAs for the strong (low $\sigma$) and weak (high $\sigma$) central target clustering present in Gaussian fields. The inset depicts a linear increase in gain with decreasing $\sigma$ that highlights the adaptability of the SA FCA to fields with an extremely dense central region.}
        \label{fig:gaussianyield}
\end{figure}

The second set consists of three groups of 40 Gaussian fields each. Each field contains 400 targets with each set parameterised by single $\sigma$ values of 0.15, 0.2 and 0.3 in order of decreasing central target clustering (see Figure \ref{fig:fields}(d) for example). SA FCA parameters are identical to those used to configure the first set. Again we see the strong performance of the SA FCA in Figure \ref{fig:gaussianyield} with the inset depicting gains of up to 11 per cent over the `Oxford' FCA. The data show the adaptability of SA to one type of high density fields, expected to be common AAOmega fields, provided that the collision matrix is calculable.

\subsubsection{Multiple Priority Fields}
The yield of individual populations of different priorities has hitherto been neglected, despite the prevalent use of such fields for combined science programs. Such yields give the clearest indication of the efficacy of FCA priority weighting regimes. One set of 500 fields with 100 uniformly distributed targets of each priority P1--9 and 100 uniformly distributed sky targets per field were generated. The fields were configured both with and without sky targets allocated. We chose $T_i$=512, equivalent to the priority weighting of P9 targets, in an attempt to incorporate each target priority in the annealing schedule.

\begin{figure}
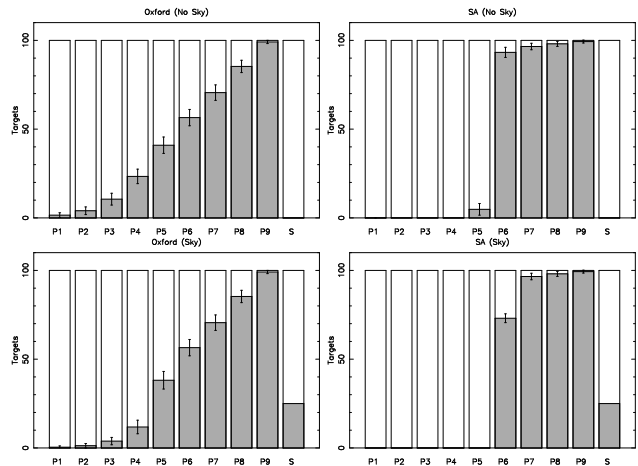

        \begin{center}
                \includegraphics[scale=0.175,angle=270]{fig8a.ps}
                \includegraphics[scale=0.175,angle=270]{fig8b.ps}\\
                \includegraphics[scale=0.175,angle=270]{fig8c.ps}
                \includegraphics[scale=0.175,angle=270]{fig8d.ps}
        \end{center}
        \caption{The average priority distributions of the fields configured by the `Oxford' (left column) and SA (right column) FCAs. Note how the `Oxford' FCA reduces the yield of P4 targets when sky targets are allocated (bottom left) despite the availability of lower priority targets. The SA FCA handles sky targets with surgical precision (bottom right), maintaining an optimal priority distribution.}
        \label{fig:priyield}
\end{figure}

Figure \ref{fig:priyield} shows the average priority distributions of the configured fields. The `Oxford' distribution is roughly linear with many low priority targets allocated. This arises from the design of the `Oxford' FCA that systematically allocates highest priority targets first. The limited number of swaps made result in minimal higher priority targets being allocated, leaving the `territory' of high priority targets open to `infiltration' by low priority targets. The SA FCA achieves a much more optimal priority distribution, with yield gains of up to $\sim$30 per cent for high priority targets in some cases, by virtue of its weighting regime and the high degree of field randomisation. 

If sky targets are included we can see the difference between the two methods of sky target allocation. If the reserved fibres cannot meet the sky target quota, then the `Oxford' FCA attempts to deallocate low priority targets in a post-configuration routine to assign fibres to sky targets. This often does not work as an already configured field resists change in a `frozen' state with very few degrees of freedom. As a result, higher priority targets are selected at the expense of the optimality of the priority distribution. The inclusion of sky targets in the annealing schedule of the SA FCA results in their allocation with such precision that the optimality of the priority distribution is not degraded. If additional weighting is given to sky targets by the SA FCA in the objective function, to meet the sky target quota, then the priority distribution may be slightly degraded but still weighted properly overall, unlike the `Oxford' FCA. 

\subsection{Uniformity}
\subsubsection{Completeness}
An important tool in unveiling the sampling behaviour of FCAs is the fraction of observed targets, or completeness, as a function of field plate position. We define the completeness as 

\begin{equation}
        C = \left\{ \begin{array}{ll}
		\displaystyle \frac{n_a}{n_a + n_u} & n_t \leq n_\mathrm{max}\\ 	
		\displaystyle \frac{n_a}{n_a + n_u}\frac{n_t}{n_{\mathrm{max}}} & n_t > n_\mathrm{max},\\ 	
	\end{array}\right.
\end{equation}
where $n_a$ and $n_u$ are the number of allocated and unallocated targets respectively, $n_t$ is the total number of targets in the distribution of interest (e.g. 500 for 500 P9 targets) and $n_\mathrm{max}$ is the maximum number of science fibres available for allocation (e.g. 392 in the absence of broken fibres). A completeness of 1.0 indicates 100 per cent coverage of a region of the field plate, whereas values of 0.7 and 1.3 represent a negative and positive bias of 30 per cent respectively. 

\begin{figure*}
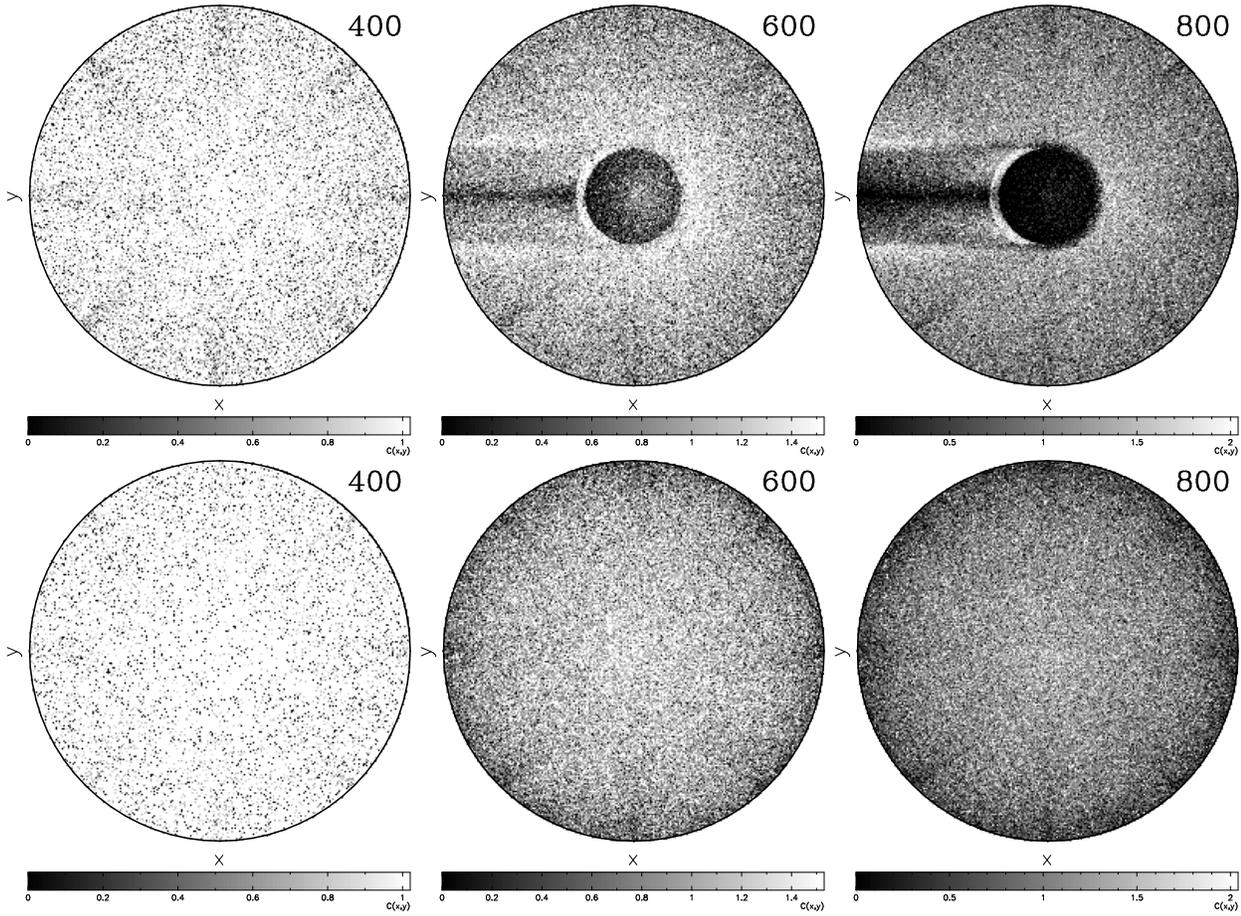

        \begin{center}
                \includegraphics[scale=0.40,angle=270]{fig9a.ps}
                \includegraphics[scale=0.40,angle=270]{fig9b.ps}
                \includegraphics[scale=0.40,angle=270]{fig9c.ps}\\
                \includegraphics[scale=0.40,angle=270]{fig9d.ps}
                \includegraphics[scale=0.40,angle=270]{fig9e.ps}
                \includegraphics[scale=0.40,angle=270]{fig9f.ps}\\
        \end{center}
        \caption{$C(x,y)$ for the `Oxford' (top) and SA (bottom) FCAs for fields including 400, 600 and 800 uniformly distributed program targets and an additional nominal 30 fiducial targets.
        The `Oxford' FCA imprints a circular deficiency and `curtain-like' feature that becomes more prominent with increasing target density. Also visible is a `flower-like' imprint by the 8 guide fibres. The high degree of randomisation used by the SA FCA eliminates much of this structure, save for a slight gradient in the radial direction. 
        }
        \label{fig:cxy}
\end{figure*}

\begin{figure*}
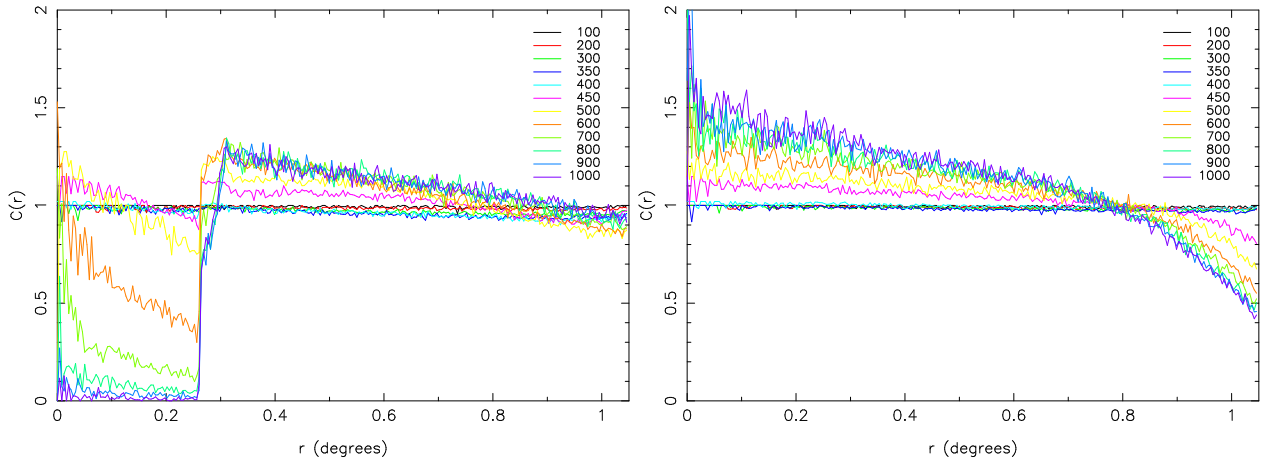

        \begin{center}
                \includegraphics[scale=0.35,angle=270]{fig10a.ps}
                \includegraphics[scale=0.35,angle=270]{fig10b.ps}
        \end{center}
        \caption{The radial completeness $C(r)$ for the `Oxford' (left) and SA (right) FCAs for fields including uniformly distributed program targets (number indicated in legend) and no fiducial targets. Notice the much higher completeness for \texttt{NTargets} $\la$ 400 produced by the SA FCA. The deficiency for $r$ $>$ 0\fdg8 imprinted by the SA FCA becomes important when all targets are of the same priority, though it is less of a concern than the circular deficiency imprinted by the `Oxford' FCA.}
        \label{fig:cruni}
\end{figure*}

Figure \ref{fig:cxy} shows $C(x,y)$, the completeness over the whole field plate, for fields containing uniform program target distributions configured by the `Oxford' and SA FCAs. Each field also includes a nominal amount of 30 fiducial targets. We increased \texttt{MaxIters} to 30 and chose a moderate value of $T_i$=100 for the SA FCA. The $n_a(x,y)$ and $n_u(x,y)$ quantities are obtained by binning allocated and unallocated target positions from hundreds of fields in a rectangular grid of resolution 30 arcsec. Alternatively, the completeness may be averaged over $\theta$ to yield the radial completeness $C(r)$ (Figure \ref{fig:cruni}).

The majority of the artificial structure evident in Figures \ref{fig:cxy} and \ref{fig:cruni} arises from the approach each FCA takes to the strong instrumental bias depicted in Figure \ref{fig:reach}. A common manifestation of the bias is seen in a radial gradient in the completeness. Attempts to `flatten' $C(r)$ by an empirically determined radial constraint in the objective function of the SA FCA were occasionally successful but would often strongly imprint artificial structure onto target distributions (MIS05). Only for $\tau \gg 1$ does the gradient become problematic in the SA FCA. 

Of more concern is the central deficiency in completeness imprinted by the `Oxford' FCA. It arises from the heuristic based upon Figure \ref{fig:reach} that is used to weight targets for allocation. It leaves those targets `easiest' to allocate ($r$ $<$ 0\fdg25) until last, allocating instead the `hardest' targets ($r$ $>$ 0\fdg25) first. The approach breaks down when $\tau \ga 1$, whereby the hardest targets fill the limited fibre quota leaving the large central deficiency seen in Figures \ref{fig:cxy} and \ref{fig:cruni}. The `curtain-like' feature seen imprinted by the `Oxford' FCA is believed to be an angular effect (OUT04). Experimental removal of this heuristic results in unpredictable behaviour (e.g. more P6 targets allocated than P9 targets, see MIS05).

Further structure is symptomatic of insufficient field randomisation. 
This is evident in the `flower-like' imprint by the 8 guide fibres on target distributions. Relatively static guide fibres may `block' other potential fibres from allocating nearby targets, resulting in lower completeness within the sector traced out by the guide fibre. For fields where $\tau \gg 1$, the effect of guide fibres on $C(r)$ is negligible. However, fields where $\tau$ $\la$ 1 inherently resist heavy randomisation and the inclusion of fiducial targets can reduce $C(r)$ for the field by as much as 2 per cent in the `Oxford' case. The high degree of randomisation introduced by the SA FCA achieves dramatic reduction in this structure.

\subsubsection{Target Priorities}
\label{sec:targetpristructure}
\begin{figure*}
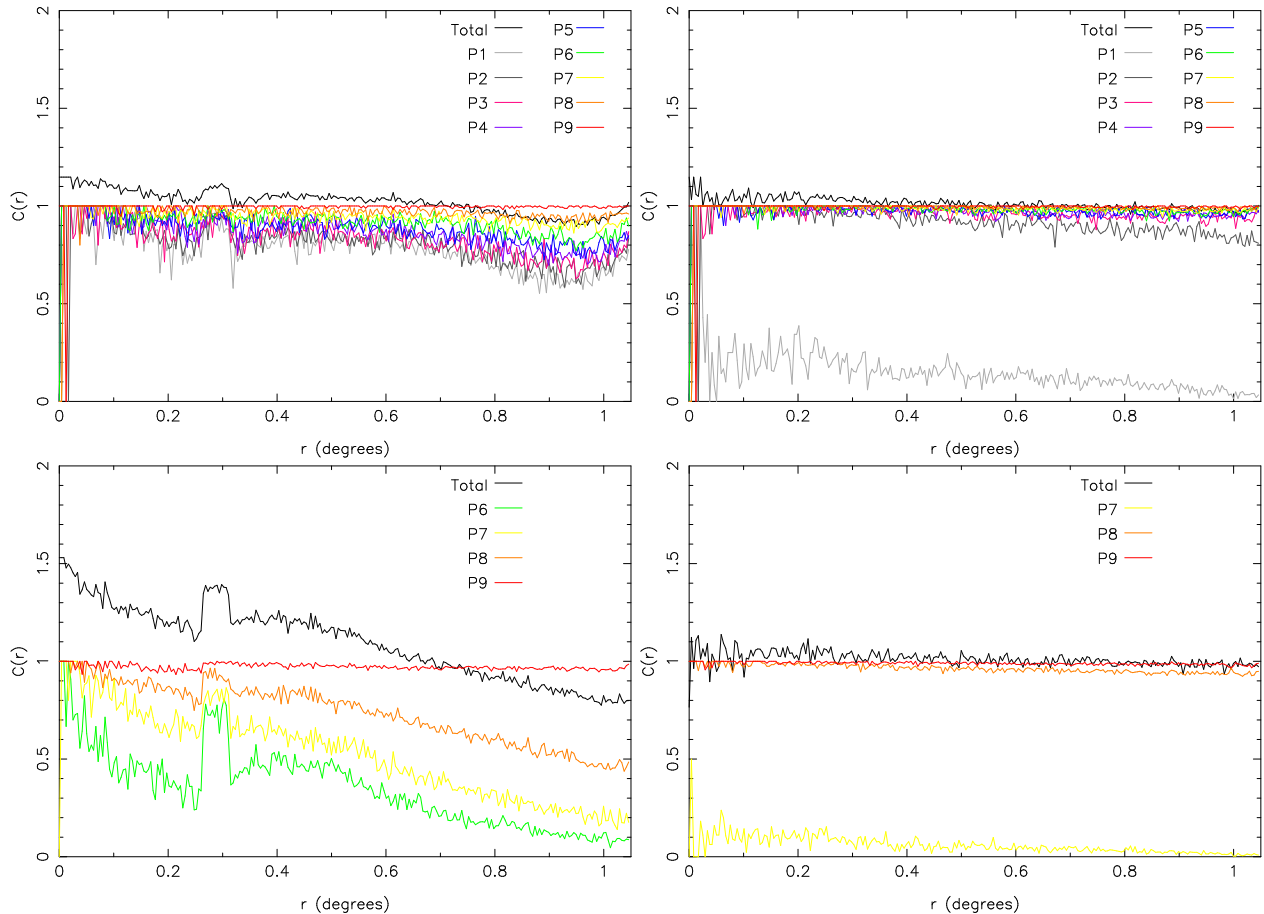

        \begin{center}
                \includegraphics[scale=0.35,angle=270]{fig11a.ps}
                \includegraphics[scale=0.35,angle=270]{fig11b.ps}\\
                \includegraphics[scale=0.35,angle=270]{fig11c.ps}
                \includegraphics[scale=0.35,angle=270]{fig11d.ps}
        \end{center}
        \caption{The radial completeness $C(r)$ for Set A (top) and Set B (bottom) fields configured by the `Oxford' (left column) and SA (right column) FCAs. Notice the strong presence of imprinted power at $r$$\sim$0\fdg25 and other gradients even at low target densities (Set A). The SA FCA completely resolves the problem by producing a very uniform response. } 
        \label{fig:cr}
\end{figure*}

Figure \ref{fig:cr} depicts the radial completeness for two sets of fields. One set (Set A) contains 50 uniformly distributed targets of each priority and the other (Set B) contains 200 P9, 200 P8, 100 P7 and 100 P6 uniformly distributed targets. The data show a startling abundance of artificial structure that the `Oxford' FCA can imprint on individual priority target populations. 

This alarming and previously unseen structure could have severe implications for surveys of low completeness that use multiple priorities in their observation strategies. For instance, the 2dFGRS assigns QSO targets higher priority in an attempt to reduce the imprint of the strong galaxy clustering signature on them. Ironically, the inclusion of lower priority galaxies meant they were made more susceptible to the imprint of artificial structure by the `Oxford' FCA. Fortunately, the high completeness of the 2dFGRS has meant that such structure is very difficult to detect, if not absent, in the final survey. This is supported by the analysis of COL01, which we have partially verified using the field and fibre information in the 2dFGRS database. Ideally, a detailed analysis should be performed with the actual configured fields for each 2dFGRS observation, however these were not archived as such. As long as target distributions of different priorities do not exceed \texttt{NPiv} in number, then sampling will generally be uniform by the SA FCA.

\subsubsection{Two-Point Angular Correlation Function}
A suitable statistic to use to quantify the angular selection function of FCAs is the two-point angular correlation function $w(\theta)$. We use the minimum variance estimator of Landy \& Szalay (1993) which is
\begin{equation}
        w(\theta) = \frac{DD - 2DR + RR}{RR}
        \label{eqn:LS}
\end{equation}
where $DD$, $DR$ and $RR$ are the number of galaxy-galaxy, galaxy-random and random-random pairs counted at the angular separation $\theta$ $\pm$ $\Delta \theta/2$, where $\log \Delta \theta$=0.1 and the number of random targets is 20 times that of program targets. We calculate the error in $w(\theta)$ as the standard deviation of the mean value of $w(\theta)$ from many different fields with the same field generation parameters.

\begin{figure}
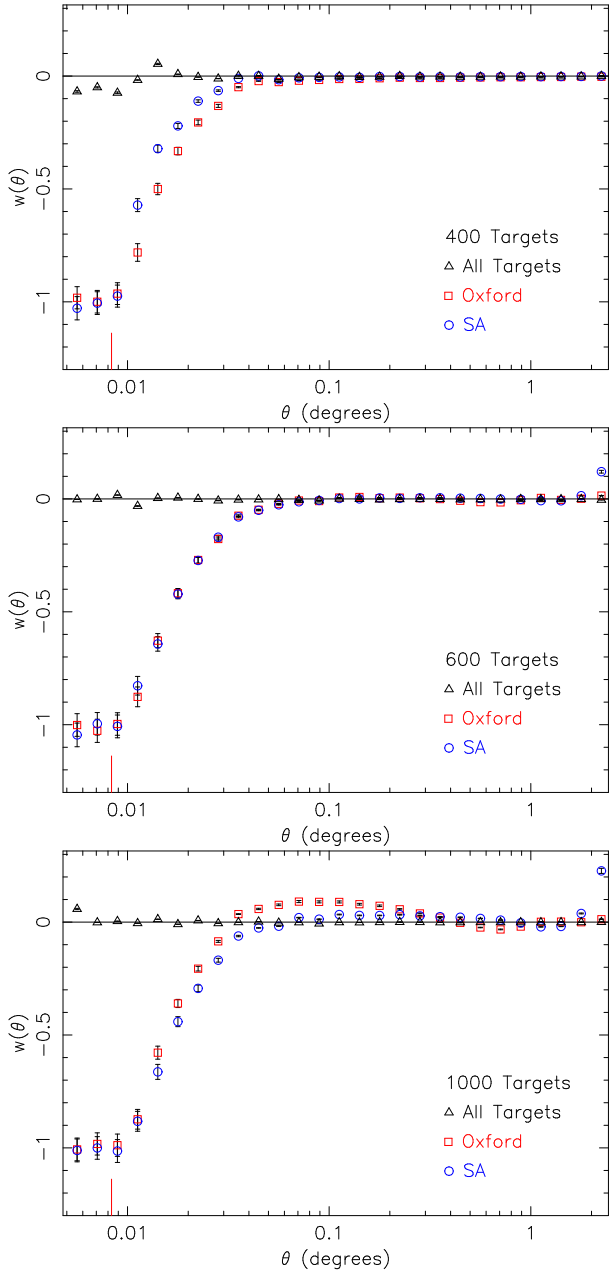

        \begin{center}
                \includegraphics[scale=0.35,angle=270]{fig12a.ps}
                \includegraphics[scale=0.35,angle=270]{fig12b.ps}
                \includegraphics[scale=0.35,angle=270]{fig12c.ps}
        \end{center}
        \caption{The average angular correlation function for fields with uniform target distributions configured by the SA and `Oxford' FCAs. The small vertical mark indicates the minimum button separation of 30 arcsec. Notice the enhanced sampling of close pairs by the SA FCA (top) and the imprint of significant structure by the `Oxford' FCA (bottom). The upturn in the SA data series at high target densities is suspected to be caused by an integral constraint.}
        \label{fig:wtheta_unif}
\end{figure}

Application to uniform fields (Figure \ref{fig:wtheta_unif}) unveils the instrumental signature convolved with the angular selection function of the FCA. The most prominent feature in the data is the strong anti-correlation in the vicinity of 30 arcsec, the minimum separation between buttons, as indicated by the small vertical mark in each graph. Other instrumental constraints such as button-fibre clearances give rise to the gradual anti-correlation up to $\sim$2.5 arcmin. For $\tau \la 1$, we see that the SA FCA naturally recovers more close pairs and both FCAs sample large angular separations uniformly. The extended hump in $w(\theta)$ for $\tau \gg 1$ imprinted by the `Oxford' FCA arises from the central deficiency depicted in Figure \ref{fig:cxy}. 

\subsubsection{Tiling Implications}
It is certainly possible that the artificial structure now evident in previous FCAs could accrue over large angular scales in surveys. One such alarming example is that shown in Figure \ref{fig:cr} that could affect surveys that use different priority populations in fields. Provided that such structure is known, sufficient steps can be taken in the survey design stage to reduce such problems. To speculate on how such structure influences galaxy correlation functions is beyond the scope of this paper. It is the responsibility of survey designers to understand the optimisation of survey uniformity on the whole, not just concerning the tiling algorithm, but incorporating a well-behaved FCA to minimise post-tiling target attrition and to ensure overall uniformity. Astronomers using the SA FCA will be reassured by the well-defined behaviour and highly uniform sampling that it provides.

\subsection{Observational Flexibility}
\subsubsection{Fibre Straightness}
Many MOS instruments have excluded fibre crossovers outright to reduce instrument complexity (e.g. Sourd 1998; Roll et al. 1998). Although crossovers can increase target yield by up to $\sim$10 per cent, a large number of crossovers is undesirable and would require all fibres to be parked before a new field was observed (LEW93). From this perspective it is desirable to minimise the number of crossovers by placing a constraint on fibre deviation to reduce field reconfiguration times. Indeed, this was first suggested by DON92 in the form of a term in the objective function for their SA based FCA. 

\begin{figure}
        \begin{center}
                \includegraphics[scale=0.35,angle=270]{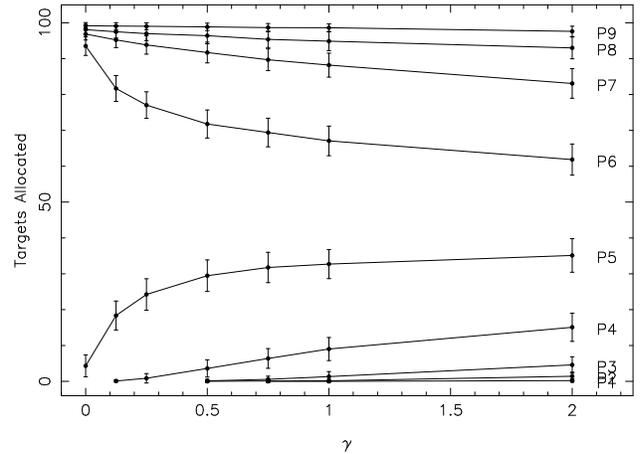}
        \end{center}
        \caption{The effect of constraining angular fibre deviation to achieve fibre straightness on the priority distribution of fields with 100 uniformly distributed targets of each priority. Note the heavy penalty on the optimality of the field with increasing $\gamma$.}
        \label{fig:fibstraight}
\end{figure}

Here we examine the effect of the angular fibre deviation constraint $\left((\alpha_i - \alpha_\mathrm{max})/\alpha_\mathrm{max}\right)^\gamma$ in Equation \ref{eqn:objective} on the priority distribution within a field. A subset of 100 fields from the set used in Figure \ref{fig:priyield} were configured by the SA FCA with $T_i$=10 fixed and varying $\gamma$. The constraint has the effect of narrowing the parameter space that each fibre can explore, resulting in progressively less optimal priority distributions with increasing $\gamma$ (Figure \ref{fig:fibstraight}). The penalty is quite severe and is not recommended for use unless all targets within a field are the same priority. If imposed on a field of single priority targets, the constraint may also reduce target yield, for example we found a deficit of $\sim$10 targets in Gaussian fields of Figure \ref{fig:gaussianyield} for $\gamma$=2. 

\subsubsection{Close Pairs}
To determine the efficacy of the close pairs constraint $\delta \sum_j^\mathrm{NAssoc} \beta^{p_j}$ in Equation \ref{eqn:objective}, 84 fields from the 2dF Mock catalogues were selected with an average number of 870 targets per field. The high density fields were chosen to ensure an abundance of close pairs. A value of $\delta$=1 was chosen with $\delta$=0.5 or 2.0 producing similar results. Close targets are identified automatically within an angular distance $\theta$, where $\theta_a \leq \theta \leq \theta_b$, and $\theta_a$ and $\theta_b$ are user-specified parameters in arcsec. 

\begin{figure}
        \begin{center}
                \includegraphics[scale=0.35,angle=270]{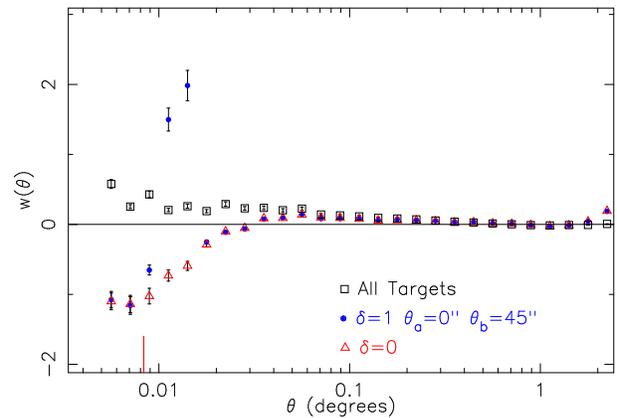}
        \end{center}
        \caption{An example of the effect of the close pairs constraint on the average $w(\theta)$ from 84 high density fields. Note the strong, isolated correlation for pairs separated by less than 45 arcsec.}
        \label{fig:closepair}
\end{figure}

A strong bias towards targets separated by $\theta \leq \theta_b$ is evident in Figure \ref{fig:closepair}. The figure shows how effective the simple constraint can be in allocating more close pairs, an otherwise very difficult task for previous FCAs. Caution is advised in using this constraint as it can strongly imprint power on observed target distributions (e.g. MIS05). 

\subsubsection{Nod \& Shuffle with Cross-Beam Switching}
Fibre orientation plays an important part in maximising allocated CBS pair yield. The footprint of optical fibres and their buttons on the field plate limits pair allocation to a select few cases where the orientation of the respective components is optimal. Target pairs with a separation $\epsilon$ oriented perpendicular to fibres have a greater chance of having both members allocated.

Here we explore the use of the close pairs constraint to maximise CBS pair yield. We generated 10 fields of 400 uniformly distributed targets that were later processed to add a pair for each target that is offset by either 30, 45 or 60 arcsec in one of two different orientations. A value of $T_i$=512 was chosen to lengthen the annealing schedule to attempt to increase the chances of more pair allocations. The `Oxford' FCA used in these tests is not optimised for CBS pair maximisation. 

\begin{figure}
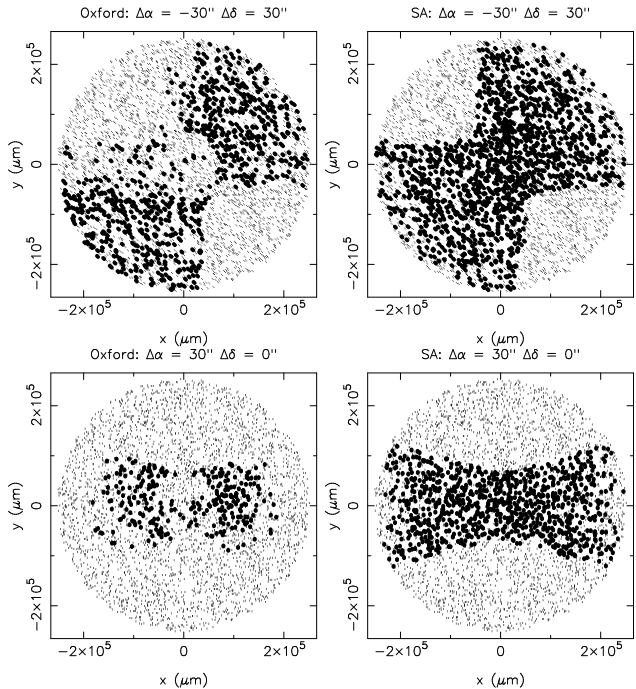

        \begin{center}
                \includegraphics[scale=0.255,angle=270]{fig15a.ps}
                \includegraphics[scale=0.255,angle=270]{fig15b.ps}\\
                \includegraphics[scale=0.255,angle=270]{fig15c.ps}
                \includegraphics[scale=0.255,angle=270]{fig15d.ps}
        \end{center}
        \caption{The distribution of allocated CBS pairs (filled circles) from 10 fields with target separations of $\epsilon$=30 arcsec in two different orientations. The `bow-tie' shape arises from the orientation of pairs to fibres. The difficulty of the problem is highlighted by the low, uneven coverage of the `Oxford' FCA. The SA FCA produces more even coverage of CBS pairs, with total coverage achieved for $\epsilon \ga 60$ arcsec.}
        \label{fig:cbsxy}
\end{figure}

\begin{figure}
        \begin{center}
                \includegraphics[scale=0.35,angle=270]{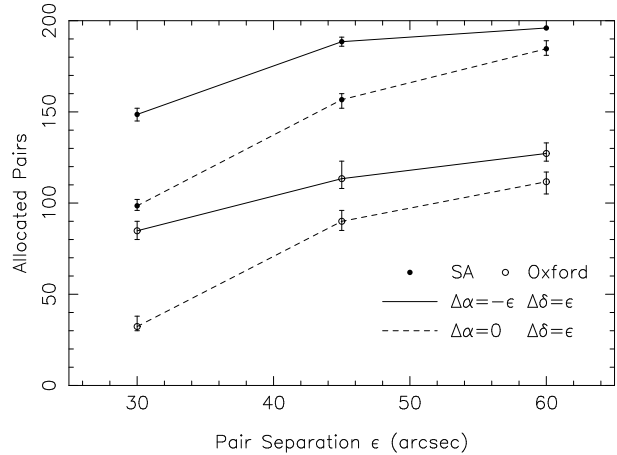}
        \end{center}
        \caption{The average number of allocated CBS pairs for fields configured by the SA (filled circles) and `Oxford' (open circles) FCAs. Error bars indicate minimum and maximum values encountered in each 10 field set described in the text. At least 98 pairs is obtained by the SA FCA even in the most difficult low $\epsilon$ limit. This is more than adequate for the 98 CBS pair limit imposed by each AAOmega CCD.}
        \label{fig:cbsyield}
\end{figure}

Figure \ref{fig:cbsxy} shows the distribution of allocated CBS pairs from fields with pairs separated by 30 arcsec in two different orientations. A `bow-tie' shape is clearly evident, a selection effect arising from the orientation of target pairs to fibres. The effect is most severe for the fields configured by the `Oxford' FCA and pairs separated by $\Delta \alpha$=30 arcsec and $\Delta \delta$=0 arcsec. The central deficiency is prominent as well as low pair yield near field edge because of the limited fibre swaps that occur in the `Oxford' FCA. The SA FCA achieves excellent uniform coverage of CBS pairs under the extremely difficult constraints, outperforming the `Oxford' FCA by $\sim$30 per cent in all cases tested (Figure \ref{fig:cbsyield}). Uniform coverage of the field is attained by the SA FCA when $\epsilon \ga 60$ arcsec in all orientations (e.g. MIS05).

\section{Conclusion}
A batch version of \textsc{configure} was created to facilitate the development of a new FCA based on SA. We bypassed preexisting \textsc{configure} collision detection strategies by pre-calculating all possible conflicts between all possible allocations and storing them in an indexed collision matrix. This allowed for field randomisation orders of magnitude greater than previously possible. Field randomisation is guided by the Metropolis algorithm with the aid of a simple objective function that can wield great influence over the final solution obtained by the FCA. Our results were obtained by configuring vast quantities of synthetic fields tailored to address specific optimality criteria. They affirm the superior quality of the new FCA that consistently satisfies the criteria to a greater extent than the previous default `Oxford' FCA for 2dF.

The new SA FCA has the following valuable attributes compared to the `Oxford' FCA:
\begin{itemize}
        \item Gains of up to 7 per cent for low target density fields.
        \item Gains of up to 11 per cent for heavily clustered Gaussian fields.
        \item Optimal target priority weighting scheme that achieves maximum yields for highest priority targets (gains of up to 30 per cent).
              
        \item Integrated sky target allocation, eliminating the unnecessary sacrifice of higher priority targets to fulfil sky target quotas.
        \item Elimination of previous artificial structure imprinted on target distributions, including the severe structure imprinted on different priority populations, to achieve highly uniform target sampling. 
        \item Quantification of sampling behaviour via the two-point angular correlation function and the completenesses $C(x,y)$ and $C(r)$.
        \item Ability to maximise the yield of close target pairs for possible use in surveys and CBS observations.
        \item Greatly improved capacity for algorithm maintenance and enhancement arising from its simple design involving the key role of the objective function.
                
\end{itemize}
The outstanding performance and flexibility of the new FCA, coupled with its generic design, makes it perfectly suitable for existing (e.g. 2dF and 6dF) and future MOS instruments with potentially little modification. Algorithm design and components such as the allocation sub-system should scale well to support future instruments with thousands of fibres, provided sufficient growth in computer memory and processing power takes place. The analytical techniques used here to analyse the imprint of artificial structure on FCAs and their general performance is recommended for use in any future FCA development. 

\section*{Acknowledgments}
BM acknowledges the support of an AAO/Macquarie University Honours scholarship to conduct this work. The authors wish to thank Phil Outram for his foresight and assistance, and Rob Sharp for useful discussions and some challenging fields to configure. We acknowledge the contribution of the `Oxford' FCA to the highly successful 2dFGRS with many thousands of observations to its credit. Only recently has modern computer hardware been capable of better field configuration. We thank the reviewer for comments that greatly improved this paper.

\end{document}